\documentclass[cameraready]{Interspeech}
\usepackage{multirow}
\usepackage{hyperref}

\title{FineCombo-TTS: Collaborative and Precise Controllable Speech Synthesis Using Text Descriptions and Reference Speech}

\address{$^1$ Shenzhen International Graduate School, Tsinghua University, Shenzhen, China \\
$^2$ Inner Mongolia University
Hohhot, China\\
$^3$ Tencent, Shenzhen, China}

\author[affiliation={1}, equalcontribution]{Shuoyi}{Zhou}
\author[affiliation={1}, equalcontribution]{Yixuan}{Zhou}
\author[affiliation={3}]{Peiji}{Yang}
\author[affiliation={2}]{Yifan}{Hu}
\author[affiliation={3}]{Yicheng}{Zhong} 
\author[affiliation={3}]{Zhisheng}{Wang}
\author[affiliation={1}, correspondingauthor]{Zhiyong}{Wu}

\email{zhousy23@mails.tsinghua.edu.cn}

\keywords{text-to-speech, controllable TTS, conditional flow matching}

\begin{document}

\maketitle

\begin{abstract}
Controllable text-to-speech (TTS) has become a key research focus. However, methods based on either reference speech or text descriptions lack flexibility and precise control, and recent joint approaches remain loosely coupled, with speech modeling timbre and text controlling global style.
We propose FineCombo-TTS, a unified framework for speech synthesis grounded in reference speech and guided by text descriptions, enabling flexible and precise control over acoustic attributes. Instead of explicit attribute disentanglement, we learn a unified acoustic representation and introduce a Conditional Flow Matching (CFM)-based Speech Variance Predictor to model fine-grained reference-to-target transformations guided by text descriptions.
To support relative attribute control, we construct FineEdit, a structured paired dataset that explicitly encodes source-to-target attribute variations. Experiments demonstrate that our approach achieves flexible, precise, and expressive controllable TTS.\footnote{The demo and dataset are available at \url{https://thuhcsi.github.io/interspeech2026-FineCombo-TTS}.}
\end{abstract}

\section{Introduction}
\label{sec:intro}

In recent years, text-to-speech (TTS) technology has achieved near human-level naturalness and intelligibility~\cite{ren2020fastspeech2, ren2019fastspeech, wang2017tacotron}, shifting research toward higher-level goals such as expressive and controllable synthesis. Existing controllable methods can be divided into three categories: 1) reference speech-based, 2) text description-based, and 3) joint speech-description control. Reference speech-based approaches~\cite{wang2023neural, kharitonov2023speartts, lei2024improving} enable zero-shot voice cloning but are highly dependent on the quality of the reference speech, limiting practical flexibility. Text description-based methods~\cite{guo2023prompttts, yang2024instructtts, liu2023promptstyle} provide greater control freedom but often struggle to capture fine-grained acoustic nuances through textual prompts alone.
Recent joint-control methods~\cite{ji2025controlspeech,zhou2024voxinstruct, li2025flespeech} attempt to combine both modalities but still operate in a segregated manner, with reference speech typically limited to timbre and text descriptions crudely overwriting global style, resulting in limited cross-modal collaboration.
\begin{figure}[t]
\centering
\centerline{
\includegraphics[width=0.9\linewidth]{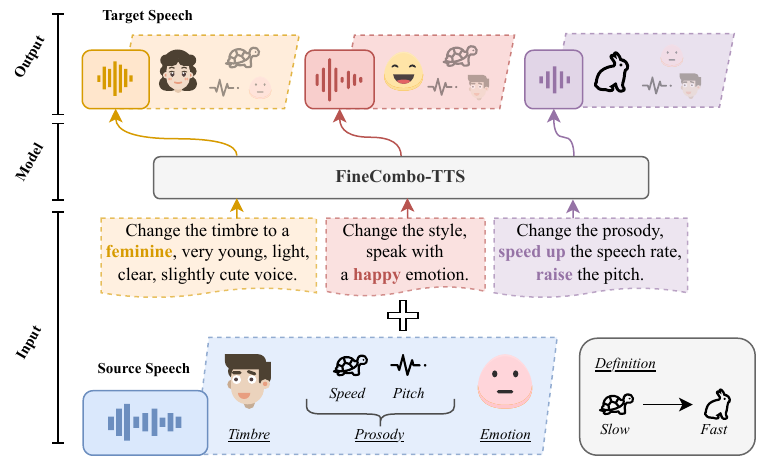}
}
\caption{The proposed FineCombo-TTS enables precise control over the timbre, prosody, and emotion of the source (reference) speech.}
\label{fig:demo}
\vspace{-13pt}
\end{figure}

Therefore, we aim to build a controllable TTS model that supports reference-aware and preference-driven speech synthesis—preserving the acoustic characteristics of a selected reference speech while enabling precise, text-guided control of specific attributes, thereby providing both precise controllability and practical flexibility.
However, achieving such joint control is limited by both data and modeling challenges. Most description-based datasets consist of isolated speech–text pairs describing absolute acoustic properties. Consequently, models learn attribute–acoustic correspondences tied to the training distribution, rather than attribute variations relative to a given reference speech. This highlights the need for structured reference–target pairs that explicitly encode relative attribute transformations. In addition, precise control is hindered by the strong entanglement among timbre, prosody, and emotion in natural speech. Explicit disentanglement is difficult and often introduces information leakage or structural redundancy, making stable and flexible control hard to achieve.

To address these issues, we construct FineEdit, a paired dataset of triplets ⟨source speech, control description, target speech⟩, where the source provides an acoustic baseline, the control description specifies the desired attribute modification, and the target reflects the corresponding variation. As shown in Fig.~\ref{fig:demo}, FineEdit covers timbre, prosody, and emotion, enabling description-guided control grounded in reference speech.
On the modeling side, we establish a unified acoustic attribute latent space, where speech attribute embeddings combine timbre information with residual style components.
We further introduce a Conditional Flow Matching (CFM)-based Speech Variance Predictor to model attribute transformations conditioned on both reference representations and text descriptions. This module learns a fine-grained, one-to-many mapping from reference-grounded attributes to controlled target variations, enabling controllable and high-fidelity speech synthesis without explicit disentanglement.
\begin{figure*}[ht]
\centering
\centerline{
\includegraphics[width=0.88\linewidth]{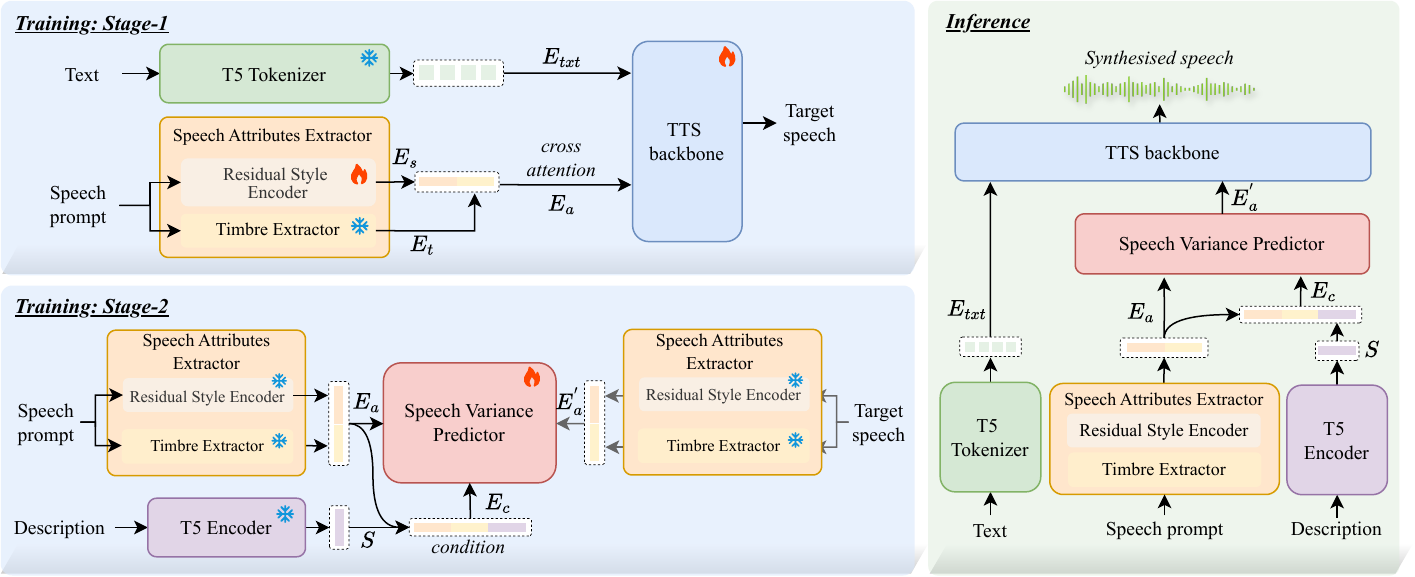}
}
\caption{The overall architecture of FineCombo-TTS.}
\label{fig:model}
\vspace{-10pt}
\end{figure*}

By integrating the structured paired dataset with our modeling framework, the proposed approach achieves reference-grounded and preference-driven controllable speech synthesis. In particular, our model supports joint control with both reference speech and text descriptions, as well as single-condition scenarios, enabling zero-shot voice cloning and description-driven speech generation. The main contributions are summarized as follows:
\begin{itemize}
\item We propose \textbf{FineCombo-TTS}, the first controllable TTS architecture that jointly leverages reference speech and text descriptions to enable flexible and precise control.
\item We introduce a CFM-based Speech Variance Predictor that operates in a unified acoustic attribute latent space, modeling fine-grained reference-to-target attribute transformations conditioned on text descriptions.
\item We construct \textbf{FineEdit}, the first large-scale paired dataset of ⟨source speech, control description, target speech⟩ triplets for relative attribute control.  Experiments validate the effectiveness of both the dataset and the proposed framework.
\end{itemize}

\section{Methodology}


The architecture of our proposed model is illustrated in Figue~\ref{fig:model}.
The model comprises three modules: a Speech Attributes Extractor, a Speech Variance Predictor, and a TTS backbone. The Speech Attributes Extractor encodes prompt speech into a unified embedding $E_a$ via a timbre extractor and residual style encoder, capturing key acoustic characteristics. The CFM-based Speech Variance Predictor refines $E_a$ conditioned on both the text encoding $S$ from T5 Encoder\cite{chung2024t5}\footnote{\href{https://huggingface.co/google/flan-t5-small}{google/flan-t5-small}} 
 and the source attributes. Finally, the TTS backbone takes text embeddings $E_{txt}$ and the target attribute $E_a'$ to autoregressively generate multi-layer acoustic tokens, which are reconstructed using the Descript Audio Codec (DAC\cite{kumar2023dac}).

\subsection{Speech Attributes Extraction}
To avoid the difficulty of explicit attribute disentanglement, we retain the natural coupling of acoustic factors and extract a unified speech attribute embedding. Specifically, we adopt the timbre extractor from pretrained FACodec\footnote{\href{https://github.com/lifeiteng/naturalspeech3_facodec}{github.com/lifeiteng/naturalspeech3\_facodec}}\cite{ju2024naturalspeech} to obtain a speaker-related timbre embedding $E_t$, leveraging large-scale pretraining for robust timbre representation. To further capture detailed style information, we adopt a residual style encoder based on the Mel-Style Encoder in~\cite{min2021metastylespeech}, which combines convolutional layers and self-attention to model both local and global patterns in the mel-spectrogram, producing a residual style embedding $E_s$. The final unified attribute embedding is constructed by concatenating the two components $E_a = concat(E_t, E_s)$.

\subsection{Speech Variance Modeling with CFM}
We design a Speech Variance Predictor that conditions on both text descriptions and reference speech, enabling precise attribute transformation without explicit disentanglement. Inspired by conditional generative frameworks such as diffusion and flow matching used in image editing\cite{hu2024imagefm, kawar2023imagic, hu2023motion}, we adopt Conditional Flow Matching (CFM)\cite{lipman2022flow} to model reference-to-target attribute variations.
Text descriptions are encoded by a pretrained T5 model, and a cross-attention module derives the sentence-level representation $S$. The predictor then conditions on the concatenation of description semantic representation $S$ and the source speech attribute embedding $E_a$.

During training, we use the source speech attribute embedding $E_a$ as $x_0$ and the target embedding $E_a'$ as $x_1$ from the FineEdit paired dataset, which differ only in the specified acoustic attribute.
For tasks that require only description control without reference speech, the source speech attribute embedding is replaced with a random noise. Linear interpolation generates intermediate states:
\begin{equation}
x_t = tx_1 + (1-t)x_0,\quad u_t = x_1 - x_0,
\label{eq45}
\end{equation}
where $t\in[0,1]$ and $u_t$ denotes the target velocity. A UNet backbone estimates the velocity field $v_t$, conditioned on $E_c=(E_a,S)$:
\begin{equation}
v_t = V_t(x_t, t \mid E_c).
\label{eq6}
\end{equation}
The training objective minimizes the MSE between $v_t$ and $u_t$:
\begin{equation}
\mathcal{L}_{CFM} = \mathbb{E}_{t,x_0,x_1}|v_t-u_t|^2.
\label{eq7}
\end{equation}

Inspired by the success of Classifier-Free Guidance (CFG)\cite{ho2022classifier} in speech generation models such as VoxInstruct \cite{zhou2024voxinstruct}, we incorporate CFG in Speech Variance Predictor to amplify the influence of text descriptions on attribute refinement. During training, we randomly drop the text condition $S$ with a fixed probability, yielding the null-conditioned embedding $E_c'=[E_a,\emptyset]$. At inference, the guidance scale $\alpha$ is employed to strengthen the text-driven transformation of speech attributes:
\begin{equation}
\hat{V_t}(x_t,t\mid E_c) = \alpha V_t(x_t,t\mid E_c) + (1-\alpha)V_t(x_t,t\mid E_c'),
\label{eq10}
\end{equation}

\subsection{TTS Backbone}
Inspired by LM-based TTS models, we adopt a decoder-only Transformer codec language model as the TTS backbone. The input text is tokenized into $E_{txt}$ and prepended as conditioning tokens. The speech attribute embedding $E_a$ is injected through cross-attention in each Transformer block, with text features as queries and $E_a$ as keys and values for attribute-aware modulation.
Following MusicGen~\cite{copet2023musicgen} and ParlerTTS~\cite{lyth2024parlertts}, we employ a delay pattern to jointly predict multi-layer acoustic tokens in a fully autoregressive manner, which facilitates coherent prosody modeling across layers and time steps. The model generates acoustic tokens with probability $P(A|E_{txt}, E_a;\theta_{TTS})$.
To improve text–speech alignment and reduce omissions or repetitions, we apply classifier-free guidance (CFG) on the text condition. During training, $E_{txt}$ is randomly dropped. At inference, a guidance scale $\beta$ strengthens adherence to the textual content:
\begin{equation}
\resizebox{1\hsize}{!}{$
\log \hat{P}(A\mid E_{txt}, E_a) = \beta \log P(A\mid E_{txt},E_a) +(1-\beta)\log P(A\mid \emptyset, E_a).
$}
\label{eq9}
\vspace{-10pt}
\end{equation}
\subsection{Training Strategy}
Our model is trained in two stages (Fig.~\ref{fig:model}). In the first stage, the FACodec timbre extractor is frozen, while the residual style encoder and TTS backbone are jointly trained on large-scale text–speech datasets, followed by fine-tuning on a smaller emotional dataset to improve expressiveness. This stage establishes stable speech generation and learns a robust unified speech attribute embedding for zero-shot scenarios. 
In the second stage, the Speech Variance Predictor is trained separately on the paired dataset, where text descriptions guide reference-to-target attribute transformations to enable precise and controllable synthesis.
\vspace{-5pt}
\section{Dataset: FineEdit}
Existing speech–description datasets provide only isolated speech–text pairs describing absolute acoustic properties, which are insufficient for learning reference-conditioned attribute transformations. To address this limitation, we construct \textbf{FineEdit}, a structured paired English speech dataset designed to model relative attribute variations.
Each sample forms a triplet $\langle$source speech, control description, target speech$\rangle$, where source and target differ in only one attribute—prosody, emotion, or timbre—while other factors remain consistent. The control description explicitly specifies the relative change. Table~\ref{table1} summarizes the three subsets. 
\subsection{Prosody Paired Speech Subset}
For prosody control, we ensure identical speaker identity and text content while varying only prosodic attributes (speed and pitch). Based on LibriTTS-R~\cite{koizumi2023libritts}, we generate controlled prosody variants using FFmpeg by adjusting speed and pitch parameters. Each utterance yields multiple versions (original, pitch-high/low, speed-fast/slow), and all pairwise combinations are constructed as source–target pairs. Each pair is annotated with a prosody-change description specifying the relative adjustment (e.g., "Change the prosody, speed up the speech rate, raise the pitch."). 
\subsection{Emotion Paired Speech Subset}
For emotion control, speaker identity is fixed while emotional expression varies. We use the Emotional Speech Database (ESD)~\cite{zhou2022esd}, which contains 10 speakers across five emotions (Happy, Sad, Angry, Surprised, Neutral). For each speaker, utterances with different emotions are paired to form source–target combinations, where the textual content is not required to be identical, and each pair is annotated with a target-style description (e.g., “Change the style, speak with an angry emotion.”).
\subsection{Timbre Paired Speech Subset}
For timbre control, we construct cross-speaker pairs while keeping overall prosody and style as consistent as possible. The target speaker description serves as the control annotation prefixed with “Change the timbre”. 
To increase data diversity and ensure that timbre variations remain stable across different affective conditions, we divide the dataset into emotion-neutral and emotion-rich subsets:

\noindent \textbf{Emotion-neutral subset}: 
Built from LibriTTS-R~\cite{koizumi2023libritts}, whose audiobook style ensures relatively consistent emotion, we group utterances with similar prosody(e.g. pitch-high, speed-fast) using speaker and prosody annotations from LibriTTS-P\cite{kawamura2024librittsp} and create cross-speaker pairs. The target speaker description serves as the control text for timbre transfer.

\noindent \textbf{Emotion-rich subset}: 
Using ESD~\cite{zhou2022esd} with prosody and emotion annotations from TextrolSpeech~\cite{ji2024textrolspeech}, we group utterances with similar prosody and emotion (e.g. pitch-high, speed-fast, emotion-happy) and construct cross-speaker pairs. We additionally provide manual speaker descriptions to support more expressive timbre control.

\begin{table}[htbp]
\centering
\caption{Summary of Attribute Control in FineEdit Paired Speech Subsets. $\checkmark$ means the target attribute is modified; $-$ means it remains unchanged.}
\resizebox{1\linewidth}{!}{
\begin{tabular}{lcccccc}
    \toprule
      \textbf{Dataset Type} & \textbf{Prosody} &\textbf{Emotion} &\textbf{Timbre} &\textbf{Text} &\textbf{Num of pairs}        \\
    \midrule
    Prosody Pair  & \checkmark   & -   & - & - &   634,956    \\
    Emotion Pair  & -   & \checkmark   & -   & \checkmark  & 80,000,000    \\
    Timbre Pair  & -   & -   & \checkmark   & \checkmark  & 16,392,828 \\
    \bottomrule
\end{tabular}
}
\label{table1}
\end{table}
\vspace{-10pt}

\section{Experiment}
\begin{table*}[!htbp]
\centering
\caption{Experimental results of prosody control.}
\resizebox{0.88\linewidth}{!}{
    \begin{tabular}{lcccccccc}
    \toprule
     \multirow{2}{*}{\textbf{Model}}& 
    \multirow{2}{*}{\textbf{MOS-S$\uparrow$}} & 
    \multirow{2}{*}{\textbf{MOS-I$\uparrow$}} & 
    \multirow{2}{*}{\textbf{WER$\downarrow$}} & 
    \multirow{2}{*}{\textbf{SECS$\uparrow$}} & 
    \multicolumn{2}{c}{\textbf{Uncontrolled Variation$\downarrow$}} &
    \multicolumn{2}{c}{\textbf{Controlled Accuracy$\uparrow$}} \\
    \cmidrule(lr){6-7} \cmidrule(lr){8-9}
    & & & & & \textbf{Speed} & \textbf{Pitch}  & \textbf{Speed} & \textbf{Pitch}  \\  
        \midrule
        VoxInstruct-Joint & 2.00 $\pm$ 0.38 & 3.26 $\pm$ 0.37 & 11.12 & 56.79 & 19.00  & 42.81  &  91.35 & 63.81   \\
        FineCombo-TTS & \textbf{4.04 $\pm$ 0.34} & \textbf{4.05 $\pm $0.31} & 12.87 & \textbf{70.20} & \textbf{14.62} & \textbf{6.71} & \textbf{98.00} & \textbf{93.33}  \\
    \bottomrule
    \end{tabular}
}
\label{table2}
\end{table*}

\begin{table*}[]
    \centering
    \caption{Experimental results of emotion control and timbre control. Emotion-A means emotion accuracy; Emotion-S means the similarity between emotion embeddings.}

    \resizebox{1\linewidth}{!}{
    \begin{tabular}{@{}lccccc|ccccc@{}}
    \toprule
    \multirow{2}{*}{\textbf{Model}} 
    & \multicolumn{5}{c|}{\textbf{Emotion Control}} & \multicolumn{5}{c}{\textbf{Timbre Control}} \\
    \cmidrule(lr){2-6} \cmidrule(lr){7-11}
  & \textbf{MOS-S$\uparrow$}  & \textbf{MOS-I$\uparrow$} & \textbf{WER$\downarrow$} & \textbf{SECS$\uparrow$} & \textbf{Emotion-A$\uparrow$}
 & \textbf{MOS-P$\uparrow$} & \textbf{MOS-I$\uparrow$} & \textbf{WER$\downarrow$} & \textbf{FPC$\uparrow$} & \textbf{Emotion-S$\uparrow$} \\
    \midrule
    VoxInstruct-Joint & 2.64 $\pm$ 0.24 & 2.96 $\pm$ 0.34 & 20.18 & 63.99 & 47.00 & 3.04 $\pm$ 0.36 & 3.32 $\pm$ 0.32 & 19.24 & 47.46 & 52.15 \\
    FineCombo-TTS & \textbf{3.34 $\pm$ 0.36} & \textbf{3.83 $\pm$ 0.18} & \textbf{11.22} & \textbf{66.56} & \textbf{85.00} & \textbf{3.66 $\pm$ 0.32} & \textbf{3.75 $\pm$ 0.27} & \textbf{18.59} & \textbf{52.67} & \textbf{55.38} \\
    \bottomrule
    \end{tabular}
    }
    \vspace{-10pt}
    \label{table3}
\end{table*}
\subsection{Experiment Setup}
In training stage 1, we pre-train on Multilingual LibriSpeech (MLS, 45k hours)~\cite{pratap2020mls} and LibriTTS-R (585 hours)~\cite{kearns2014librivox}, and then fine-tune on EmoVoice-DB (45 hours)~\cite{yang2025emovoice} and TextrolSpeech (330 hours)~\cite{ji2024textrolspeech} for emotion modeling. In stage 2, we train the Speech Variance Predictor using 236K description–speech pairs from TextrolSpeech for description-only control, together with about 600K pairs per FineEdit subset for precise joint control with both reference speech and text.

The TTS backbone is a 12-layer Transformer decoder, and the CFM framework employs a 1D UNet as the predictor. To enable unconditional CFG training, we randomly mask text or description sequences with probability 0.1, and set the CFG strength to $\alpha=\beta=2$ during inference. 
All experiments are conducted on 8×NVIDIA A100 GPUs. In stage 1, we train for 250K steps with a batch size of 32 and a learning rate of $1\text{e}^{-4}$, followed by fine-tuning on emotional datasets for 70K steps with a learning rate of $5\text{e}^{-4}$. In stage 2, the Speech Variance Predictor is trained for 140K steps with a learning rate of $1\text{e}^{-4}$.

Since no prior model achieves joint processing of reference speech and text descriptions, we adapt an existing open-source SOTA framework for fair comparison. Specifically, we re-implement VoxInstruct~\cite{zhou2024voxinstruct} as \textbf{VoxInstruct-Joint}. On top of the original fully LLM-based architecture, we modify it to support joint control by prepending the acoustic tokens of the reference speech to the input sequence. To ensure fairness, VoxInstruct-Joint is trained on the same data as our method, including FineEdit, under an identical training strategy.
\vspace{-3pt}
\subsection{Experiment Results}
We evaluate controllability on prosody, emotion, and timbre using both reference speech and textual descriptions. Test data are sampled from FineEdit and unseen during training. 
Evaluation combines subjective metrics—\textbf{MOS-S} (speaker similarity), \textbf{MOS-I} (instruction following), and \textbf{MOS-P} (prosodic consistency)—with objective metrics: \textbf{WER}\footnote{\href{https://huggingface.co/openai/whisper-large-v3}{openai/whisper-large-v3}}  for intelligibility, \textbf{SECS}\footnote{\href{https://huggingface.co/microsoft/wavlm-base-plus-sv}{microsoft/wavlm-base-plus-sv}} for speaker encoder cosine similarity, \textbf{FPC} for pitch correlation, and \textbf{Emotion-A/S}\footnote{\href{https://huggingface.co/emotion2vec/emotion2vec_plus_large}{emotion2vec/emotion2vec\_plus\_large}} for emotion accuracy and embedding similarity. For joint-control analysis, we also report \textbf{Controlled Accuracy}, which measures whether the synthesized attribute change matches the instruction, and \textbf{Uncontrolled Variation}, which quantifies how much non-target attributes deviate from the reference (lower is better).

For prosody control (Table~\ref{table2}), FineCombo-TTS achieves higher MOS-I and Controlled Accuracy than VoxInstruct, confirming stronger instruction following. Meanwhile, MOS-S and SECS remain high, indicating that timbre is preserved, and Uncontrolled Variation in speed and pitch is much lower, showing that specific attributes can be modified precisely without degrading others.

For emotion control (Table~\ref{table3}, left), FineCombo-TTS reaches an emotion accuracy of 85\%, significantly outperforming VoxInstruct. At the same time, high MOS-S and SECS demonstrate that the speaker’s timbre remains well preserved during emotion transfer.

For timbre control (Table~\ref{table3}, right), FineCombo-TTS achieves MOS-I of 3.75 for timbre instructions and surpasses the baseline in MOS-P, FPC, and Emotion-S. This demonstrates effective timbre modification while maintaining prosody and emotional style, validating the benefit of CFM-based conditional control.
\subsection{Ablation Studies}
\noindent \textbf{Different CFG Strategy}
We evaluate the CFG strategy on the emotion control task with multi-CFG with text CFG value of 2.0 and description CFG value of 1.5. As shown in Table~\ref{table6}, strengthening description CFG improves instruction following, raising emotion accuracy from 81\% to 86\%, though SECS decreases slightly since stronger emotion expression can reduce timbre similarity. Moreover, adding text CFG lowers WER from 14.17 to 9.06, enhancing intelligibility and naturalness. Overall, multi-CFG achieves better instruction adherence while maintaining a good balance between speaker similarity and naturalness.
\vspace{-5pt}
\begin{table}[h]
    \centering
    \caption{Ablation study on different CFG strategy.}
    \resizebox{1\linewidth}{!}{
    \begin{tabular}{lccc}
    \toprule
        Model & WER$\downarrow$ & SECS$\uparrow$ & Emotion-A$\uparrow$ \\
        \midrule
        w/o CFG on description and text & 14.17 & \underline{71.08} & 76.00\\
        w/o CFG on description & \underline{9.06}  & \textbf{72.53} & \underline{81.00}  \\
        proposed & \textbf{8.82} & 69.16 & \textbf{86.00} \\
        \bottomrule
    \end{tabular}
    }
    \label{table6}
\end{table}
\vspace{-5pt}

\noindent \textbf{Ablation of Residual Style Encoder}
In our model, we adopt the pre-trained FACodec timbre extractor for speaker information and introduce a residual style encoder for detailed style modeling. We assess the contribution of the residual style encoder in the first-stage zero-shot experiments. As shown in Table \ref{table7}, incorporating the residual style encoder leads to improvements in both speaker similarity and speech quality, indicating its positive contribution to modeling speaker style.
\vspace{-5pt}
\begin{table}[h]
    \caption{Ablation study of residual style encoder.}
    \centering
    \begin{tabular}{lcc}
    \toprule
        Model & MCD$\downarrow$ & SECS$\uparrow$ \\
        \midrule
        w/o residual style encoder& 11.08 & 90.00\\
        proposed & \textbf{10.83} & \textbf{90.20}\\
        \bottomrule
    \end{tabular}
    \label{table7}
\end{table}
\vspace{-10pt}
\section{Conclusion}

In this paper, we propose FineCombo-TTS, a controllable TTS framework that tightly integrates reference speech and text descriptions for precise and flexible speech synthesis. Unlike prior pseudo-collaborative approaches, our method unifies both modalities within a shared acoustic attribute space, enabling reference-grounded and text-guided attribute transformation.
We introduce a CFM-based Speech Variance Predictor to model fine-grained reference-to-target attribute variations without explicit disentanglement.
To support such relative control learning, we also construct FineEdit, a structured paired dataset that explicitly encodes attribute differences between source and target speech.

\section{Generative AI Use Disclosure}
During the preparation of this manuscript, the authors used generative AI tools exclusively for the purpose of language editing and manuscript polishing to improve readability. These tools were not used to generate any core scientific ideas, experimental data, or technical contributions. All authors have thoroughly reviewed and approved the final version of the manuscript, and assume full responsibility for the integrity and entirety of its content.

\section{Acknowledgments}
This work was supported by National Natural Science Foundation of China (62076144) and National Social Science Foundation of China (13\&ZD189).

\bibliographystyle{IEEEtran}
\bibliography{mybib}

@article{wang2023neural,
  title={Neural codec language models are zero-shot text to speech synthesizers},
  author={Wang, Chengyi and Chen, Sanyuan and Wu, Yu and Zhang, Ziqiang and Zhou, Long and Liu, Shujie and Chen, Zhuo and Liu, Yanqing and Wang, Huaming and Li, Jinyu and others},
  journal={arXiv preprint arXiv:2301.02111},
  year={2023}
}

@article{kharitonov2023speartts,
  title={Speak, read and prompt: High-fidelity text-to-speech with minimal supervision},
  author={Kharitonov, Eugene and Vincent, Damien and Borsos, Zal{\'a}n and Marinier, Rapha{\"e}l and Girgin, Sertan and Pietquin, Olivier and Sharifi, Matt and Tagliasacchi, Marco and Zeghidour, Neil},
  journal={Transactions of the Association for Computational Linguistics},
  pages={1703--1718},
  year={2023},
  publisher={MIT Press One Broadway, 12th Floor, Cambridge, Massachusetts 02142, USA~…}
}

@inproceedings{lei2024improving,
  title={Improving language model-based zero-shot text-to-speech synthesis with multi-scale acoustic prompts},
  author={Lei, Shun and Zhou, Yixuan and Chen, Liyang and Luo, Dan and Wu, Zhiyong and Wu, Xixin and Kang, Shiyin and Jiang, Tao and Zhou, Yahui and Han, Yuxing and others},
  booktitle={ICASSP 2024-2024 IEEE International Conference on Acoustics, Speech and Signal Processing (ICASSP)},
  pages={12662--12666},
  year={2024},
  organization={IEEE}
}

@inproceedings{guo2023prompttts,
  title={Prompttts: Controllable text-to-speech with text descriptions},
  author={Guo, Zhifang and Leng, Yichong and Wu, Yihan and Zhao, Sheng and Tan, Xu},
  booktitle={ICASSP 2023-2023 IEEE International Conference on Acoustics, Speech and Signal Processing (ICASSP)},
  pages={1--5},
  year={2023},
  organization={IEEE}
}

@article{yang2024instructtts,
  title={Instructtts: Modelling expressive tts in discrete latent space with natural language style prompt},
  author={Yang, Dongchao and Liu, Songxiang and Huang, Rongjie and Weng, Chao and Meng, Helen},
  journal={IEEE/ACM Transactions on Audio, Speech, and Language Processing},
  pages={2913--2925},
  year={2024},
  publisher={IEEE}
}

@article{liu2023promptstyle,
  title={Promptstyle: Controllable style transfer for text-to-speech with natural language descriptions},
  author={Liu, Guanghou and Zhang, Yongmao and Lei, Yi and Chen, Yunlin and Wang, Rui and Li, Zhifei and Xie, Lei},
  journal={arXiv preprint arXiv:2305.19522},
  year={2023}
}

@article{li2025flespeech,
  title={Flespeech: Flexibly controllable speech generation with various prompts},
  author={Li, Hanzhao and Li, Yuke and Wang, Xinsheng and Hu, Jingbin and Xie, Qicong and Yang, Shan and Xie, Lei},
  journal={arXiv preprint arXiv:2501.04644},
  year={2025}
}

@inproceedings{ji2025controlspeech,
  title={ControlSpeech: Towards Simultaneous and Independent Zero-shot Speaker Cloning and Zero-shot Language Style Control},
  author={Ji, Shengpeng and Chen, Qian and Wang, Wen and Zuo, Jialong and Fang, Minghui and Jiang, Ziyue and Huang, Hai and Wang, Zehan and Cheng, Xize and Zheng, Siqi and others},
  booktitle={the 63rd Annual Meeting of the Association for Computational Linguistics (Volume 1: Long Papers)},
  pages={6966--6981},
  year={2025}
}

@inproceedings{zhou2024voxinstruct,
  title={Voxinstruct: Expressive human instruction-to-speech generation with unified multilingual codec language modelling},
  author={Zhou, Yixuan and Qin, Xiaoyu and Jin, Zeyu and Zhou, Shuoyi and Lei, Shun and Zhou, Songtao and Wu, Zhiyong and Jia, Jia},
  booktitle={the 32nd ACM International Conference on Multimedia},
  pages={554--563},
  year={2024}
}

@article{ju2024naturalspeech,
  title={Naturalspeech 3: Zero-shot speech synthesis with factorized codec and diffusion models},
  author={Ju, Zeqian and Wang, Yuancheng and Shen, Kai and Tan, Xu and Xin, Detai and Yang, Dongchao and Liu, Yanqing and Leng, Yichong and Song, Kaitao and Tang, Siliang and others},
  journal={arXiv preprint arXiv:2403.03100},
  year={2024}
}

@article{lipman2022flow,
  title={Flow matching for generative modeling},
  author={Lipman, Yaron and Chen, Ricky TQ and Ben-Hamu, Heli and Nickel, Maximilian and Le, Matt},
  journal={arXiv preprint arXiv:2210.02747},
  year={2022}
}

@article{ho2022classifier,
  title={Classifier-free diffusion guidance},
  author={Ho, Jonathan and Salimans, Tim},
  journal={arXiv preprint arXiv:2207.12598},
  year={2022}
}

@inproceedings{hu2024imagefm,
  title={Latent space editing in transformer-based flow matching},
  author={Hu, Vincent Tao and Zhang, Wei and Tang, Meng and Mettes, Pascal and Zhao, Deli and Snoek, Cees},
  booktitle={AAAI conference on artificial intelligence},
  pages={2247--2255},
  year={2024}
}

@inproceedings{kawar2023imagic,
  title={Imagic: Text-based real image editing with diffusion models},
  author={Kawar, Bahjat and Zada, Shiran and Lang, Oran and Tov, Omer and Chang, Huiwen and Dekel, Tali and Mosseri, Inbar and Irani, Michal},
  booktitle={IEEE/CVF conference on computer vision and pattern recognition},
  pages={6007--6017},
  year={2023}
}

@article{hu2023motion,
  title={Motion flow matching for human motion synthesis and editing},
  author={Hu, Vincent Tao and Yin, Wenzhe and Ma, Pingchuan and Chen, Yunlu and Fernando, Basura and Asano, Yuki M and Gavves, Efstratios and Mettes, Pascal and Ommer, Bjorn and Snoek, Cees GM},
  journal={arXiv preprint arXiv:2312.08895},
  year={2023}
}

@inproceedings{ji2024textrolspeech,
  title={Textrolspeech: A text style control speech corpus with codec language text-to-speech models},
  author={Ji, Shengpeng and Zuo, Jialong and Fang, Minghui and Jiang, Ziyue and Chen, Feiyang and Duan, Xinyu and Huai, Baoxing and Zhao, Zhou},
  booktitle={ICASSP 2024-2024 IEEE International Conference on Acoustics, Speech and Signal Processing (ICASSP)},
  pages={10301--10305},
  year={2024},
  organization={IEEE}
}

@article{lyth2024parlertts,
  title={Natural language guidance of high-fidelity text-to-speech with synthetic annotations},
  author={Lyth, Dan and King, Simon},
  journal={arXiv preprint arXiv:2402.01912},
  year={2024}
}

@article{kumar2023dac,
  title={High-fidelity audio compression with improved rvqgan},
  author={Kumar, Rithesh and Seetharaman, Prem and Luebs, Alejandro and Kumar, Ishaan and Kumar, Kundan},
  journal={Advances in Neural Information Processing Systems},
  pages={27980--27993},
  year={2023}
}

@article{chung2024t5,
  title={Scaling instruction-finetuned language models},
  author={Chung, Hyung Won and Hou, Le and Longpre, Shayne and Zoph, Barret and Tay, Yi and Fedus, William and Li, Yunxuan and Wang, Xuezhi and Dehghani, Mostafa and Brahma, Siddhartha and others},
  journal={Journal of Machine Learning Research},
  number={70},
  pages={1--53},
  year={2024}
}

@inproceedings{min2021metastylespeech,
  title={Meta-stylespeech: Multi-speaker adaptive text-to-speech generation},
  author={Min, Dongchan and Lee, Dong Bok and Yang, Eunho and Hwang, Sung Ju},
  booktitle={International Conference on Machine Learning},
  pages={7748--7759},
  year={2021},
  organization={PMLR}
}

@article{copet2023musicgen,
  title={Simple and controllable music generation},
  author={Copet, Jade and Kreuk, Felix and Gat, Itai and Remez, Tal and Kant, David and Synnaeve, Gabriel and Adi, Yossi and D{\'e}fossez, Alexandre},
  journal={Advances in Neural Information Processing Systems},
  pages={47704--47720},
  year={2023}
}

@article{ren2020fastspeech2,
  title={Fastspeech 2: Fast and high-quality end-to-end text to speech},
  author={Ren, Yi and Hu, Chenxu and Tan, Xu and Qin, Tao and Zhao, Sheng and Zhao, Zhou and Liu, Tie-Yan},
  journal={arXiv preprint arXiv:2006.04558},
  year={2020}
}

@article{ren2019fastspeech,
  title={Fastspeech: Fast, robust and controllable text to speech},
  author={Ren, Yi and Ruan, Yangjun and Tan, Xu and Qin, Tao and Zhao, Sheng and Zhao, Zhou and Liu, Tie-Yan},
  journal={Advances in neural information processing systems},
  year={2019}
}

@article{wang2017tacotron,
  title={Tacotron: Towards end-to-end speech synthesis},
  author={Wang, Yuxuan and Skerry-Ryan, RJ and Stanton, Daisy and Wu, Yonghui and Weiss, Ron J and Jaitly, Navdeep and Yang, Zongheng and Xiao, Ying and Chen, Zhifeng and Bengio, Samy and others},
  journal={arXiv preprint arXiv:1703.10135},
  year={2017}
}

@article{koizumi2023libritts,
  title={Libritts-r: A restored multi-speaker text-to-speech corpus},
  author={Koizumi, Yuma and Zen, Heiga and Karita, Shigeki and Ding, Yifan and Yatabe, Kohei and Morioka, Nobuyuki and Bacchiani, Michiel and Zhang, Yu and Han, Wei and Bapna, Ankur},
  journal={arXiv preprint arXiv:2305.18802},
  year={2023}
}

@article{zhou2022esd,
  title={Emotional voice conversion: Theory, databases and esd},
  author={Zhou, Kun and Sisman, Berrak and Liu, Rui and Li, Haizhou},
  journal={Speech Communication},
  pages={1--18},
  year={2022},
  publisher={Elsevier}
}

@article{kawamura2024librittsp,
  title={Libritts-p: A corpus with speaking style and speaker identity prompts for text-to-speech and style captioning},
  author={Kawamura, Masaya and Yamamoto, Ryuichi and Shirahata, Yuma and Hasumi, Takuya and Tachibana, Kentaro},
  journal={arXiv preprint arXiv:2406.07969},
  year={2024}
}

@article{kearns2014librivox,
  title={Librivox: Free public domain audiobooks},
  author={Kearns, Jodi},
  journal={Reference Reviews},
  pages={7--8},
  year={2014},
  publisher={Emerald Publishing}
}

@article{pratap2020mls,
  title={Mls: A large-scale multilingual dataset for speech research},
  author={Pratap, Vineel and Xu, Qiantong and Sriram, Anuroop and Synnaeve, Gabriel and Collobert, Ronan},
  journal={arXiv preprint arXiv:2012.03411},
  year={2020}
}

@article{yang2025emovoice,
  title={Emovoice: Llm-based emotional text-to-speech model with freestyle text prompting},
  author={Yang, Guanrou and Yang, Chen and Chen, Qian and Ma, Ziyang and Chen, Wenxi and Wang, Wen and Wang, Tianrui and Yang, Yifan and Niu, Zhikang and Liu, Wenrui and others},
  journal={arXiv preprint arXiv:2504.12867},
  year={2025}
}

\end{document}